\documentclass[
aps,%
11pt,%
final,%
notitlepage,%
oneside,%
onecolumn,%
nobibnotes,%
nofootinbib,% In the current version of REVTeX, when this option is on,
superscriptaddress,%
noshowpacs,%
centertags]%
{revtex4}

\RequirePackage{graphicx}

\def\refitem#1{\relax}

  \newfont{\cyrfnt}{wncyr9 scaled 1120}

\begin{document}

\title{Signals of Deconfinement Phase Transition and Possible Energy Range of Its Detection}

\author{\firstname{K.A.} \surname{Bugaev}}
\email{Bugaev@th.physik.uni-frankfurt.de}
\affiliation{Bogolyubov Institute for Theoretical Physics, NAS of Ukraine, Metrologichna str. 14$^B$, Kiev 03680, Ukraine}

\maketitle

\centerline{\bf Abstract}
{\small
Here we thoroughly discuss  the present status of  the deconfinement  phase transition signals
outlined in the NICA White Paper 10.01. It is argued that none of  the signals   outlined in the NICA White Paper
is prepared  for   experimental verification.
At the same time we discuss the new irregularities and new signals of the deconfinement
transition found recently within the realistic version of the hadron resonance gas model. All new findings evidence that the mixed quark-gluon-hadron phase 
can be reached at the center of mass energy of collision 4.3-4.9 GeV.
}

\section{Present status of  traditional deconfinement signals} 

After nearly three decades of the  heavy ion experiments  and the  searchers for the quark gluon plasma 
the situation with the experimental signals of its formation is not clear at the moment.
Although some irregularities, widely known as the Kink \cite{Kink}, the Strangeness Horn \cite{Horn} and the Step \cite{Step},  were observed and are often attributed   to 
the onset of deconfinement \cite{Gazd_rev:10}, their relation to deconfinement  was not clarified. 
It is necessary to admit that, despite  the multiple claims of the authors of Ref. \cite{Gazd_rev:10} about existence of the
statistical model of early stage, in fact, all three above  mentioned  irregularities cannot  be explained within a single framework and, hence,
no one exactly understands   what these irregularities really signal.  

An actual  difficulty with a physical interpretation of these irregularities appears due to the fact that  up to now they cannot 
be reproduced either within the transport models like UrQMD and HSD or  within the hybrid hydro-cascade models \cite{Hybrid:1}.  If  in the case of the pure transport models one can always argue that 
the failure   happens due to 
an absence of the 1-st order phase transition (PT) in these models,  this argument does not work for the 
hybrid hydro-cascade model since in such a model the hydrodynamics is used just to model a PT. 
Therefore, it is quite possible that we do not understand something very basic at the level of the equation of state with the 1-st order PT.  The second possible reason of the hybrid hydro-cascade model  failure
is that the employed  interaction between hadrons  is oversimplified.  The validity of the last statement 
was demonstrated very recently by the success of  the multicomponent resonance gas model \cite{HRGM:13} in a simultaneous description  of  the Strangeness Horn  in the ratio of  $K^+$ and $\pi^+$ multiplicities    and  a similar peak in  the ratio of  $\Lambda$  and $\pi^-$ multiplicities.  
Note that such a model  employes 
different hard-core radii for pions, kaons, all other mesons  and that one for baryons \cite{HRGM:13},
which, so far, have not been used in the transport  or hybrid  hydro-cascade models. 
Therefore, additional  and independent justification of the irregularities suggested in 
\cite{Kink,Horn, Step} is necessary. 
This task is rather important in  view of the planned heavy-ion collision  experiments at 
JINR NICA  and  GSI FAIR.

It is evident that searching for other irregularities and signals of mixed phase formation and 
their justification is no less significant. Some  efforts to suggest and to discuss new signals 
of mixed phase formation were made in the NICA White Paper \cite{NICAwp}. 
However, the present situation with the vast majority of  deconfinement and /or mixed phase formation
signals does not look optimistic. As an example let us  briefly consider a situation with a  promising signal of the chiral symmetry restoration transition, known as  the chiral magnetic effect (CME). 
Prior to  the moment, when local P and CP violations can be included into 
hydrodynamic, transport and hybrid hydro-cascade models, the situation  with the CME  physical interpretation 
as a signal of the chiral symmetry restoration transition will remain unclear.  Therefore, all the problems 
and tasks of future research on CME  discussed in the NICA White Paper  sections 8.1, 8.2, 8.3 and  8.5 remain actual in the 
coming years.  A similar statement is valid for the chiral vortaic effect discussed in the NICA White Paper section 8.6, i.e. before the P-odd effects are implemented into and studied by  hydrodynamic, transport and hybrid hydro-cascade models, it is hard to give  a physical interpretation for  the chiral vortaic effect.

Hence in this work we critically analyze the typical signals  of  the deconfinement  PT which are outlined in the NICA White Paper 10.01 \cite{NICAwp} and discuss their pitfalls.  At the same time we present here  the new irregularities and new signals of the deconfinement
transition found recently  within the realistic version of the hadron resonance gas model \cite{HRGM:13}.

\section{Phase transitions in finite systems and the concept of  spinodal phase decomposition} 

Due to an absence of  the first principle theory of  PTs in finite systems  the classical, i.e. nonstatistical,  models of phase transition become very popular.  The typical examples of such models are the usual Van der Waals  equation of state and the mean-field models. The main problem with such models is that they 
do not obey the second and third L. van Hove axioms of statistical mechanics \cite{Axioms,Axioms2}.
In particular, in the classical models a PT  occurs even at very small number of particles in the system (just a few particles). Such a behavior contradicts to the third  axiom of statistical mechanics \cite{Axioms,Axioms2} and to experimental fact that a PT is washed out in a finite system. 
The major  physical reason of  these troubles is that the classical models  have incorrect parameterization  of the gaseous phase neglecting the well known fact that in real  systems the gas  does not  consists just of  the molecules, but it  consists of the   molecule  clusters (or droplets) of all possible sizes \cite{FDM}. As a consequence, in the mean-field  models  the mechanisms  of phase transition and critical endpoint generation are not realistic and this leads to a  contradiction
with the  L. van Hove axioms of statistical mechanics \cite{Axioms,Axioms2}.
Therefore, the conclusions based on such classical models with a PT are not reliable and, hence, they should be verified with the statistical models. 

A typical example of the classical model usage  in the NICA  White Paper  is the model of  spinodal 
phase decomposition discussed in the NICA White Paper sections 3.3 and 4.19. The approach outlined in these sections 
might be interesting for further exploration, but it is dealing solely with the  Van der Waals like  equation of state. Moreover,  it explicitly assumes a concept of uniform matter which   during the evolution through the mixed phase region is transformed into the clumps under some additional assumptions and somewhat questionable approximations. 
Such an approach leads to several   principal questions which require a lot of additional work to be done.  
To be more convincing, it would be nice, if 
 the  authors of sections 3.3 and 4.19  provide some additional  theoretical arguments and corresponding numerical estimates explaining  why in a finite system the isotherms would resemble the isotherms of  Van der Waals like  equation of state found for an infinite system. 
 So far, this key assumption of sections 3.3 and 4.19 was not throughly justified. 
 
Furthermore, if  at the region of the  deconfinement  PT the quark gluon plasma behaves as the strongly interacting liquid \cite{Shuryak:sQGP}, then it is reasonable to assume that the framework of  statistical cluster models  of  liquid-gas 1-st order PT, such as  the famous Fisher droplet model 
\cite{FDM} 
of gas condensation,
the statistical multifragmentation model  \cite{SMM:1, SMM:1b,SMM:2} which describes the nuclear liquid-gas PT, and  an exactly solvable statistical model of quark gluon bags with surface tension (QGBST)  \cite{QGBST1, QGBST2},  is  better suited to model  a deconfinement PT,  than any classical equation of state. 
In these statistical models there is no uniform matter in the mixed phase region. Moreover, one does not need to invent any clumps in this case because within  these statistical  cluster models the gaseous phase already consists of the clusters of all possible sizes, which in case of  the deconfinement PT are the quark gluon plasma bags  \cite{Kapusta:81}.  Finally,  in section 4.19 there is a very interesting observation on  the spinodal amplification of density fluctuations at NICA energies. Suppose  that  an  approach of section 4.19 is absolutely correct. Then   it would be nice, if 
the authors  of   section 4.19  specify  how
one can detect this spinodal amplification of density fluctuations, because up to now we do not have  any  method to extract the spatial attenuation of the particle density from the existing experimental data.

A more elaborate approach to study the dynamical aspects of the QCD PTs is based  on the nonequilibrium chiral fluid dynamics (the NICA White Paper section 4.20). Its recent development presented in  \cite{Dynamics:13} clearly demonstrate that  a delay of a  PT  and interaction between the chiral field and quark-antiquark liquid  
leads to dissipation and noise, which in turn affect  the field fluctuations and lead to the formation  of  the net baryon density domains  \cite{Dynamics:13} which in a sense are similar to clumps.   Thus, the present approach can serve
as some, but  very preliminary, justification for the spinodal 
phase decomposition discussed in sections 3.3 and 4.19. 

However, even this elaborate approach  of the NICA White Paper section 4.20  is not consistent with the exact analytical solutions of  such statistical cluster models as  the statistical multifragmentation model  \cite{SMM:FV} and the gas of  hadronic bags \cite{GBM:FV} in finite volumes.  The authors of sections 3.2  and 4.20 assume that in a finite system the free energy has two local minima which correspond to the macroscopic pure phases, while the mixed phase corresponds to two equal values of  free energy of these minima. However, such an  assumption is based on the mean-field, i.e. classical   equations of state. At the same time, the exactly solvable  statistical cluster models  
\cite{SMM:FV, GBM:FV} tell us that (i) in a finite system the analog of liquid phase may exists at an infinite pressure only, while at finite pressures there  can exist only the finite volume analogs of gaseous and mixed phases. 
Also the  models  \cite{SMM:FV,GBM:FV} tell us that (ii)  in a finite system  the gaseous phase always consists 
 of  a single state whose free energy is real and it has the minimal value compared to  
other possible states in a finite system. These  other  states consist of  an even number of
metastable states (2, 4, 6 , ... depending on the temperature $T$ and baryonic chemical potential $\mu$)  
which come in pairs of complex conjugate  values of  free energy.  
Then the finite volume analog of a mixed phase consists of  a stable gaseous phase and an even number   of metastable sates. 
Moreover, (iii) one can rigorously show that the states  which  belong to the finite volume analog of mixed phase are not in a true chemical equilibrium with each other,  since in finite systems  the realistic interaction between hadrons and quark gluons bags differently  modifies the chemical potential  of each of these states.  
Therefore, in contrast to the approaches of  the NICA White Paper sections 3.2, 3.3,  4.19 and 4.20,   in finite systems one has 
to consider an ensemble of  a stable gaseous state and a finite even number  of metastable states. 
Furthermore, it was  shown that the imaginary part of the free energy  $I_n$ of a metastable state $n = 2, 4, 6, ...$ defines  the decay/formation time $t_n \sim 1/|I_n|$ of the $n$-th state \cite{GBM:FV}.  According to the last  formula the gaseous state  is always stable since its decay time $t_1 \rightarrow \infty$ because $I_1 = 0$. Thus, it seems that  the life time of  the $n$-th metastable state in the finite volume  analog of mixed phase cannot be taken arbitrarily, but it should depend on the imaginary part of  the $n$-th state free energy.  An important question is, however, how can one implement this property into the existing  nonequilibrium chiral fluid dynamics?

\section{Signals of the  QCD phase diagram (tri)critical endpoint} 

The situation with the (tri)critical  endpoint of the QCD phase diagram is even less clear than with the observed signals of
the onset on deconfinement. So far, it is not exactly known whether in our physical world the  QCD phase diagram  has a critical or a tricritical endpoint.  Also up to now  there is no first principal definition of the finite volume analog of  the (tri)critical  endpoint  ((tri)CEP). As a results we are forced to use  the (tri)CEP definition suited for   infinite systems.
A typical example of such an approach is outlined in the NICA White Paper section 3.1, where 
it is explicitly assumed that in a finite system the (tri)CEP has the same properties as in an infinite system, but with a finite  correlation length. Then the whole framework how to possibly detect the  (tri)CEP  in the experiment is based on such an assumption. 

Also  in the NICA White Paper section 3.1 it is implicitly assumed that the  definition of  the  correlation length used for static system is
correct for an expanding system created in heavy ion collision. 
However, according to section 4.20 the dynamical study of the correlations and fluctuations is at the very beginning. 
The usage of the static definition of correlation length in an evolving system is highly non-trivial. 
Moreover, according to section 4.20  the result for the global correlation length in a finite expanding system depends 
on the way  of the correlation length averaging. Therefore, it seems that in finite  non-static systems the (tri)CEP  definition via the correlation length is not a reliable one even for the classical model of PT and, hence,  we need something 
more reliable. 

Furthermore, the idea  to study the multiplicity and $p_T$ spectra  fluctuations ($p_T$ is the transversal momentum of secondary hadron)  on the event-by-event basis  outlined in section 3.1   does not look very optimistic at the moment.  Indeed, if the correlation length 
is finite (and small compared to the typical system size), then  very precise measurements are  necessary to 
distinguish the Gaussian and non-Gaussian shapes of the fluctuating  quantities. At the moment it is not completely clear whether the existing (and the  planned) experimental set-ups are (will be)  able to provide the necessary accuracy. Therefore,  such 
estimates are very necessary. 

In addition, there are two major difficulties with an  interpretation    of  hadron multiplicities  and/or  $p_T$ spectra  fluctuations. 
\begin{enumerate}
\item Even in a single  collision event  the kinetic freeze-out of a given hadron species occurs not at the same thermodynamic parameters, since the different space-time elements of freezing hadron gas have their own thermodynamic parameters at freeze-out. Therefore, the resulting attenuation is already spread over some (not well controlled!)  range  of the freeze-out 
temperature $T_{fo}$ and baryonic chemical potential $\mu_{fo}$. Thus, from the very beginning one has to develop a  dynamical treatment of fluctuations  in a single even of collisions. This is very tough task. 

\item The measured hadronic multiplicities and  $p_T$ spectra usually contain a sizable contribution of non-thermal 
hadrons coming from the decay of heavy resonances.  According to the NICA White Paper section 4.7 (see also Refs. [85, 86, 88] of this section)  in order to get the trustworthy results for  hadron multiplicity fluctuations it is absolutely necessary to have a very well defined input of the hadronic mass spectrum including the uncertainty in degeneracy and decay patterns of the high lying resonances, as well as the experiment specific week decays feed-downs. And  here there is a principle problem. If the quark gluon bags are, indeed, the highly lying resonances  \cite{ResModif:12}, then for a successful
analysis of fluctuations one should include into a  treatment all high  lying resonances. But then one should also know the above mentioned parameters (degeneracies, decay patterns and so on) of these highly lying resonances
which at the moment can only be guessed \cite{FWM:09}. 

Therefore, at the moment one cannot expect that   the statistical models with the truncated hadronic mass spectrum 
can provide us with  a reliable information from the experimentally measurable fluctuation patterns 
for all center of mass energies of collision  above 2  GeV.   
Of  course, one can believe that a failure of the statistical hadronization model (or similar models) in describing 
the experimentally observed fluctuations  can be an indirect signal of the quark gluon bags appearing. However, if we need a clear signal, then it will be necessary to work out more realistic, but, consequently,  more complicated statistical and transport  models which include both the hadronic and quark gluon bag mass spectra. 
\end{enumerate}

These general remarks are applicable in full  to the   net-proton number fluctuations analysis discussed in section 3.10.
In addition, an important related task is to elucidate the relation between the net-proton number fluctuations and the existing statistical models with the (tri)CEP and the 1-st order PT of a liquid-gas type \cite{QGBST1, QGBST2, FWM:09,PowerLaw}. Moreover, in order  to understand the meaning of experimentally measured skewness and kurtosis,  a similar investigation should be made for the same models, but for finite volumes. 

An interesting and important application of  the finite size scaling known from  spin models to a description of experimental data suggested  in \cite{Roy1411}  allowed one to establish the location  of the QCD phase diagram endpoint at the center of mass energy   at $\sqrt{s_{NN}} = 47.5$~GeV.  This approach also demonstrates that a universality class of the QCD phase diagram endpoint coincides with the one of the 3-dimensional Ising model  \cite{Roy1411}.  However, 
again no definite conclusion can be made whether this is a critical or a tricritical endpoint. Note that this important issue is simply ignored in the NICA White Paper \cite{NICAwp}.

\section{New Irregularities and New Signals} 

Formulation of reliable experimental signals of the deconfinement PT  was and is one of the major tasks of heavy ion  phenomenology. 
However,  until very  recently such efforts were not    
successful,  since they require the realistic equation of state and the model, statistical or dynamical,  which is  able to
accurately  describe the existing experimental data and, thus, provide us with 
reliable information about the late  stages of the heavy-ion collision process. 
Fortunately, 
the recent improvements \cite{Oliinychenko:12,HRGM:13,KAB:gs,BWGwidth} of the hadron 
resonance gas model  (HRGM)
\cite{KABCleymans:93,KABCleymansFO,PBM:99,KABpbm:02,KABAndronic:05,KABAndronic:09} provide us with the most successful description  of available hadronic multiplicities measured in heavy-ion collisions  at AGS, SPS, and RHIC energies. 
The detailed description of the HRGM and the sets of  used experimental data can be found in the original works \cite{Oliinychenko:12,HRGM:13,KAB:gs,BWGwidth}, while here we concentrate on its results. 
The global values of $\chi^2/dof \simeq 1.16$  
and  $\chi^2/dof \simeq 1.06$  achieved by the HRGM, respectively, in 
Refs. \cite{HRGM:13} and  \cite{KAB:gs}  for 
111 independent multiplicity ratios measured at 
{14} values of  the center-of-mass energy $\sqrt{s_{NN}} = $  from 2.7 GeV to 200~GeV (more details can be found in \cite{HRGM:13,KAB:gs}).
Up to date this is the best quality of the fit achieved by the comparable models.
This fact 
give us a full confidence that  the irregularities  found  in the narrow range of collision energies 
$\sqrt{s_{NN}} = 4.3-4.9$~GeV are not artifacts of the model, but, indeed, they  reflect a real situation. 

However,  the thirty years  experience of heavy ion community  shows that a formulation of a signal
 of  mixed phase formation is not the most difficult part of the problem, since 
 very many  signals were suggested, but only a few of them were observed. 
 The real trouble  to formulate a convincing model  of the  heavy-ion collision process, which 
would allow one  to connect a certain  irregularity  in 
the behavior of some observable  with  the occurrence  of  the  QGP-hadron phase transition. 
In other words,  the hardest part is to verify a suggested signal on the  existing theoretical back up and on the existing experimental data.

\begin{figure}[htbp]
  \centerline{
%%  \begin{minipage}[c]{0.49\textwidth}
    \includegraphics[height=70.7mm]{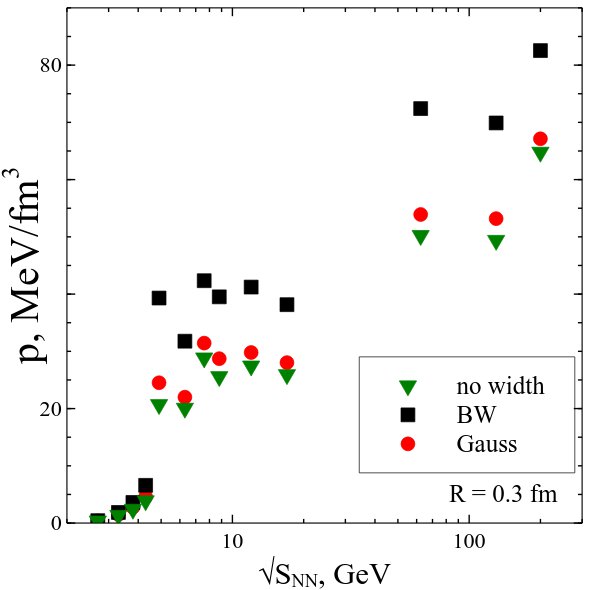}
 %% \end{minipage}
  \hspace*{1.cm}
%%  \begin{minipage}[c]{0.49\textwidth}
   \includegraphics[height=73mm]{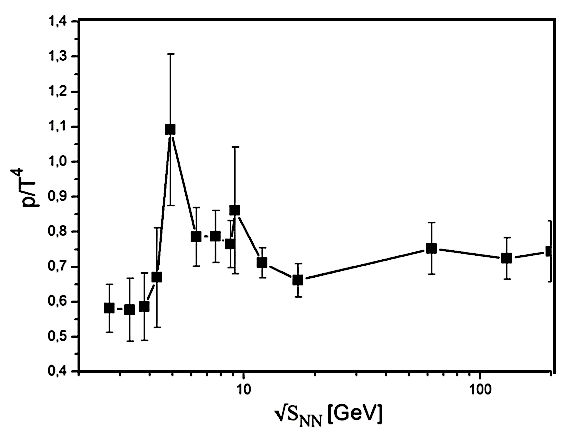}
%%  \end{minipage}
  }
 \caption{{\bf Left panel:} Comparison of the chemical FO pressures  found within the HRGM  with a single hard-core radius for all hadrons \cite{BWGwidth}, but  for different parameterizations of the  width of hadronic resonances.  The Breit-Wigner  width parameterization is given by the squares, the Gaussian width is shown by the circles, while a zero width is indicated by the triangles. The results of the HRGM with the multicomponent hard-core repulsion \cite{HRGM:13} are practically the same.
 {\bf Right  panel:} The number of effective degrees of freedom found for the HRGM with the multicomponent hard-core repulsion \cite{HRGM:13}.
 }
  \label{Fig1}
\end{figure}

Therefore, our first task was to refine the HRGM and  to convert it into a precise tool for obtaining reliable information about the stage of chemical freeze-out (FO). This aim was achieved in  Refs. \cite{Oliinychenko:12,HRGM:13,KAB:gs,BWGwidth}.  From the left panel of  Fig.\ 1 one can see
that the chemical FO pressure $p$  unprecedentedly    jumps in 4 - 7 times, depending on the  width parameterization for hadronic resonances,  if  the collision energy $ \sqrt{s_{NN}}$ changes from 4.3 GeV to 4.9 GeV.  From the right panel of Fig.\ 1 it is seen that the number of effective degrees of freedom  also experiences sizeable increase of about 70\% for the model with the multicomponent hard-core repulsion. This means that  the hard-core  radii for pions, $R_{\pi}$, kaons, $R_K$, 
other mesons, $R_m$, and baryons, $R_b$, are different. The best  global fit  of all hadronic multiplicities 
was found for $R_b$ = 0.2 fm, $R_m$ = 0.4 fm, $R_{\pi}$ = 0.1 fm, and $R_K$ = 0.38 fm   
\cite{HRGM:13}. More details about this  version of  HRGM   can be found in 
Refs. \cite{Oliinychenko:12,HRGM:13,KAB:gs}.  It is necessary to stress that all results found within a realistic version of the HRGM with a single hard-core radius of all hadrons are practically the same for the multicomponent version and vice versa, while the unrealistic versions, which correspond to an ideal gas case or the  case with hadrons of vanishing width,  usually demonstrate somewhat weaker  effects.

\begin{figure}[htbp]
  \centerline{
%%  \begin{minipage}[c]{0.49\textwid
    \includegraphics[height=70.7mm,width=77.mm]{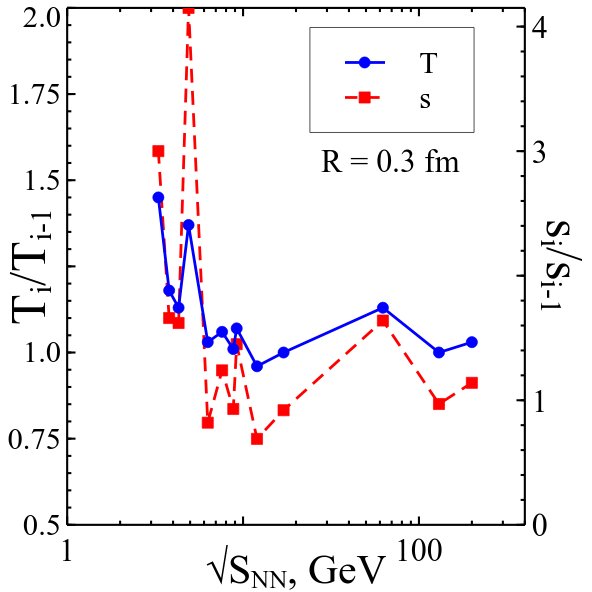}
 %% \end{minipage}
  \hspace*{1.cm}
%%  \begin{minipage}[c]{0.49\textwidth}
   \includegraphics[height=70.7mm]{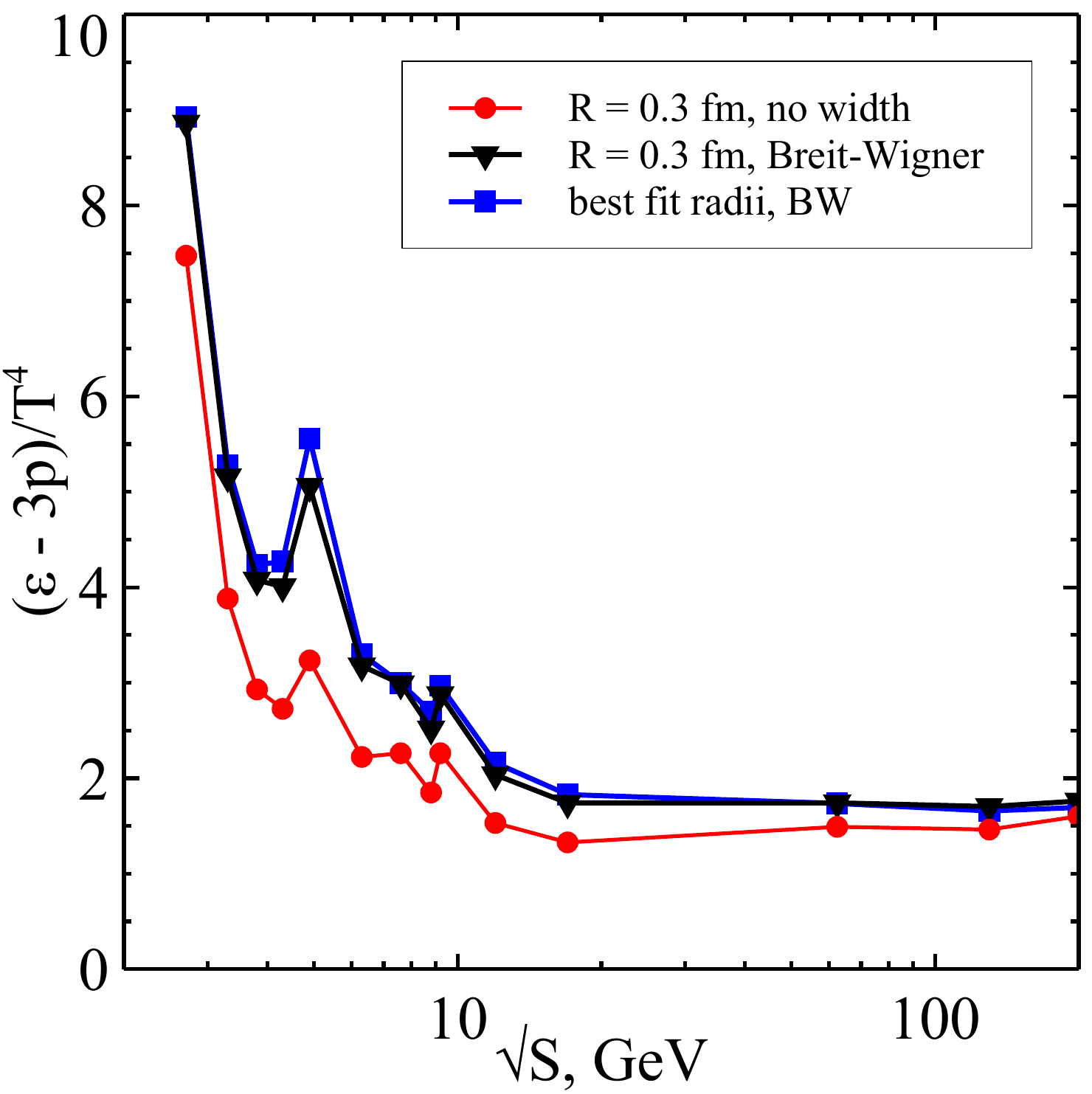}
%%  \end{minipage}
  }
 \caption{{\bf Left panel:} Ratio of chemical FO temperatures $T^{FO} (\sqrt{s_{NN}(i)}/ T^{FO} (\sqrt{s_{NN}(i-1)} )$ and ratio of entropy densities $s^{FO} (\sqrt{s_{NN}(i)}/ s^{FO} (\sqrt{s_{NN}(i-1)} )$ for two subsequent energies of collision ($i\ge 2$) are shown for the HRGM with the Breit-Wigner width parameterization. For the Gaussian  width parameterization the results are practically the same \cite{BWGwidth}. The lines connected the symbols are given to guide the eyes.
 {\bf Right  panel:} The trace anomaly at chemical FO is shown for the HRGM with a single hard core radius  \cite{BWGwidth} and for the one with the multicomponent hard-core repulsion \cite{HRGM:13}.
 }
  \label{Fig2}
\end{figure}
%%%%%%%%%%%%%%%%%%%%%%%%%%%%%

Fig.\ 2 demonstrates other peculiar  irregularities at the same energy range  $ \sqrt{s_{NN}}=$ 4.3--4.9 GeV, From the left  panel of  Fig.\ 2 it is clearly seen that in this  narrow range of collision energy the chemical FO temperature $T^{FO}$ increases in about 1.35 times,
while the entropy density at chemical FO in this case jumps in about 4.2 times! 
Another  remarkable  irregularity, the behavior of the dimensionless trace anomaly
$\delta =  \frac{\varepsilon - 3p}{T^4}$,  is shown in the right panel of  Fig. \  2.
One may 
what is the reason for  the sharp maxima shown in Fig. \ 2. 
Evidently, this irregularities are related to the behavior of the effective number of degrees of freedom, since 
\begin{eqnarray}
\delta \equiv T \frac{\partial }{\partial T} \left( \frac{p}{T^4}\right) + \mu_B  \frac{\partial }{\partial 
\mu_B} \left( \frac{p}{T^4}\right) + \mu_{I3}  \frac{\partial }{\partial \mu_{I3}} \left( \frac{p}
{T^4}\right)\,,
\end{eqnarray}
where $\mu_B$ is the baryonic chemical potential and $\mu_{I3}$ is the chemical potential of the isospin third projection. 
On the other hand,  the entropy density $s$ is a derivative of pressure with respect to temperature,
i.e. $s = \frac{\partial p}{\partial T}$. Therefore, it is not a 
coincidence that  the peak of trance anomaly and the peak of the entropy ratio $s^{FO} (\sqrt{s_{NN}(i)}/ s^{FO} (\sqrt{s_{NN}(i-1)} )$   occur  right at the energy, at which 
chemical FO pressure has a strong jump.

\begin{figure}[htbp]
  \centerline{
    \includegraphics[height=70.7mm,width=77.mm]{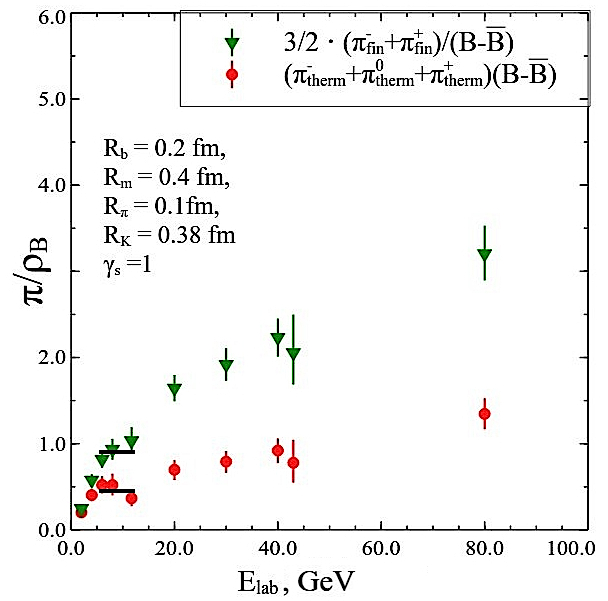}
\hspace*{11mm}
    \includegraphics[height=70.7mm,width=77.mm]{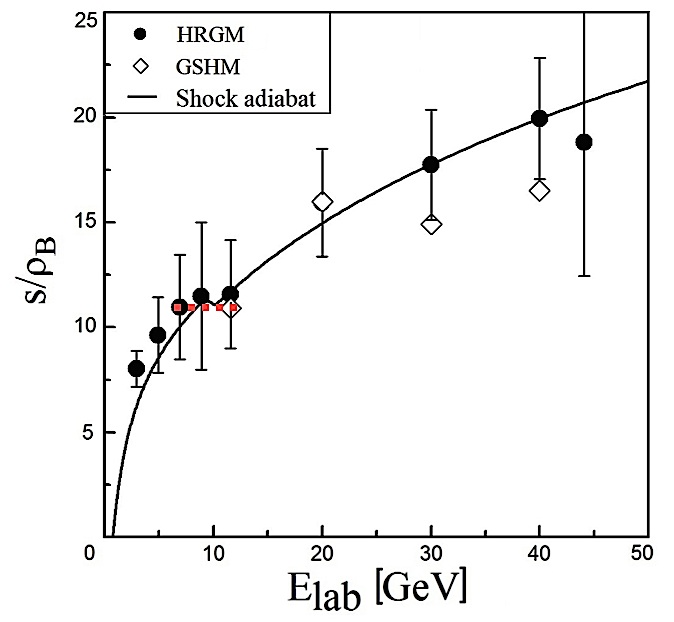}}
 \caption{ {\bf Left panel:} The energy dependence of  thermal pions per baryon (circles) and
 the charged pions per baryon (triangle) at chemical FO found for the HRGM with multicomponent repulsion \cite{HRGM:13}. The horizontal bars indicate the region of correlated plateaus found in the laboratory energy range  6.9--11.6~GeV ($\sqrt{s_{NN}} = 3.8-4.9$~GeV) \cite{SA1}.
 {\bf Right panel:} Same as in the left panel, but for the chemical FO entropy per baryon.  The
 circles correspond to the  HRGM \cite{HRGM:13}, while the diamonds  are shown for the 
 model with chemical nonequilibrium  \cite{GSHM}. The dashed horizontal line is the plato correlated with the two plateaus shown in the left panel. 
 }
  \label{Fig3}
\end{figure}

Moreover, a sharp peak in the trace anomaly $\delta$ is naturally explained within 
the shock adiabat model of central nuclear collisions \cite{SA1} as  a formation of the mixed quark-gluon-hadron phase. In Ref.  \cite{SA2} it was found 
that the trace anomaly peak at chemical FO (see Fig.\ 2)  is generated by the corresponding  trace anomaly peak on  the shock  adiabat, which appears at the boundary  between the mixed phase and  the quark gluon plasma.  Therefore,  the chemical FO peak of $\delta$ is one of the most spectacular signals of  deconfinement PT. However, one should keep in mind that for the unrealistic versions of the HRGM (i.e. without hard-core interaction or with a vanishing width of resonances) 
the $\delta$ peak  is rather weak (see the right panel of Fig.\ 2).
 
The other  signal of the deconfinement PT found in  \cite{SA1} is the set 
of  correlated  plateaus in the collision-energy dependence of the entropy 
per baryon, and
of the  total and the thermal  numbers of pions per baryon, 
which were predicted a long  time ago 
\cite{KAB:89a,KAB:90,KAB:91} to be a manifestation  of the anomalous thermodynamic properties of the mixed phase.  In fact, these plateaus  provide us, probably, with the first deconfinement PT  signal which is observed exactly as it was predicted in  \cite{KAB:89a,KAB:90,KAB:91}.
Only the energy range of this signal  estimated in \cite{KAB:89a,KAB:90,KAB:91} is lower because 
25 years ago there were no realistic equations of state which could be thoroughly verified on experimental data available at present.  It has to be stressed that the behavior of entropy per baryon found 
in other  models of chemical FO is very similar. This is seen  in the right panel of Fig. \ 3 from the comparison of  the HRGM results with the results of the entirely different model developed in \cite{GSHM}.

\begin{figure}[htbp]
  \centerline{
%%  \begin{minipage}[c]{0.49\textwidth}
   \hspace*{1.1mm} \includegraphics[height=60.7mm]{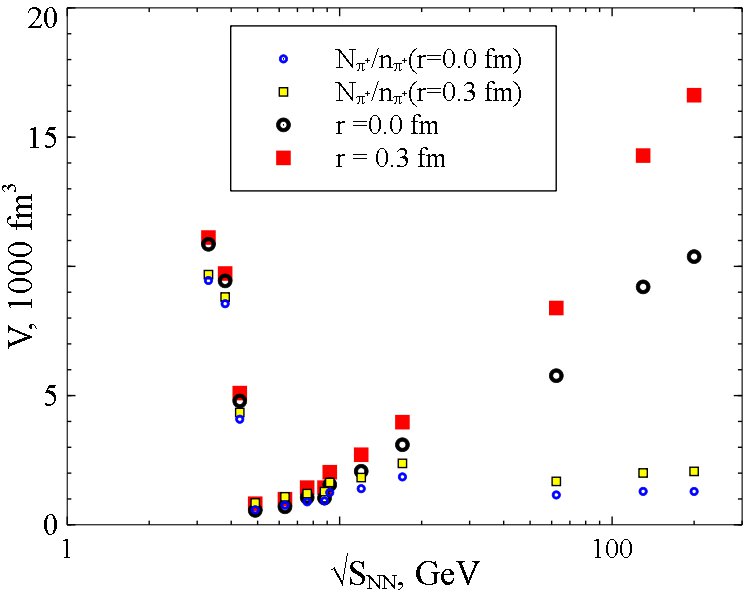}
 %% \end{minipage}
  \hspace*{-2.2mm}
%%  \begin{minipage}[c]{0.49\textwidth}
   \includegraphics[height=60.7mm]{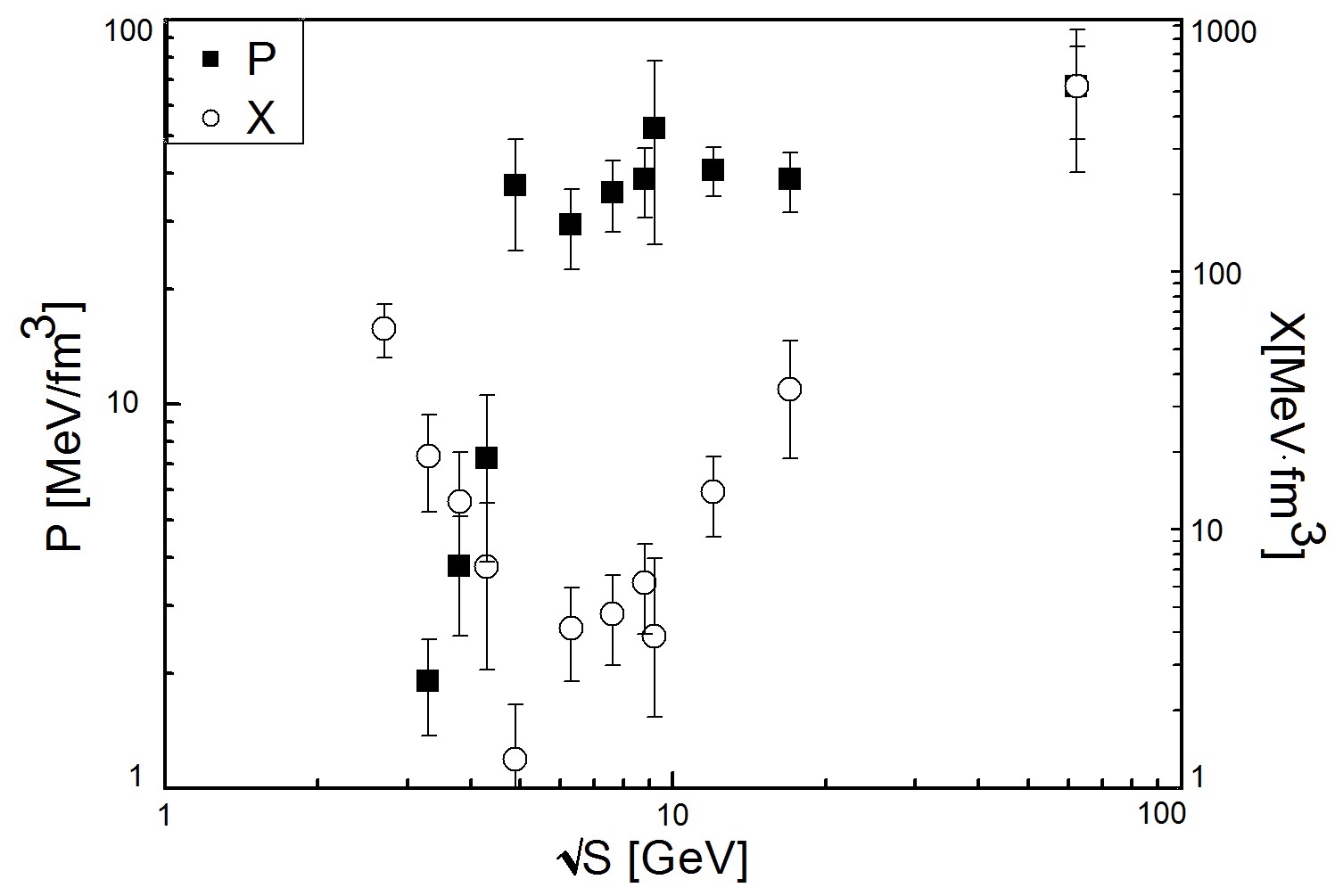}
%%  \end{minipage}
  }
 \caption{{\bf Left panel:} The chemical freeze-out volume vs. $\sqrt{s_{NN}}$ for the ideal hadron gas and the hadron gas with the same  hard core radius 0.3 fm. The smaller symbols correspond to the fit of hadron yield ratios for the nonstandard  set of  conservation laws used in 
 \cite{KABAndronic:09}, while the larger symbols are obtained by the fit of hadron multiplicities with the standard set of conservation laws  \cite{Oliinychenko:12}. In either case there is a local minimum at  $\sqrt{s_{NN}} = 4.9$~GeV (or $E_{lab} = 11.6$ GeV).
  {\bf Right  panel:} The relativistic generalized specific volume $X$ (circles) and pressure $p$ (squares) at chemical FO as the functions of   collision energy.  The  minimum of $X$ exists at   the same energy $\sqrt{s_{NN}} = 4.9$~GeV as the minimum of the chemical FO volume and jump of the chemical FO pressure.   
 } 
  \label{Fig4}
\end{figure}

The numerical simulations of the shock adiabat inside the mixed phase allow one to naturally explain the reason  why it was so difficult to pin it  down experimentally during last 30 years.  The reason is that the whole mixed phase is located within a narrow range of energy (between $\sqrt{s_{NN}} = 3.8$ and  $\sqrt{s_{NN}} = 4.9$~GeV, at most, as it is seen from Fig.\ 3), at which there is a lack of experimental data. At the same time,  a  good description of the entropy per baryon found at chemical FO within the shock adiabat model (see the solid curve in the right panel of  Fig.\ 3)  made  it possible to extract the equation of state quark gluon plasma at high baryonic chemical potentials directly from the chemical FO data \cite{SA1}. 

Furthermore, the  shock adiabat model allowed one to explain an old puzzle of the minimum 
in the chemical FO volume existing at  $\sqrt{s_{NN}} = 4.9$~GeV (see the left panel of Fig. \ 4).  In  Refs. \cite{SA1} it is shown that this minimum is directly related to the minimum 
of the relativistic generalized specific volume $X \equiv (\varepsilon + p)/ \rho_B^2$ which is defined via the pressure $p$, the energy density $\varepsilon$  and the baryinc charge density 
$\rho_B$ (see the right panel of Fig. \ 4). Moreover,  it is possible to prove \cite{SA1} that the minimum of $X$ at chemical FO is generated by the $X$ minimum along the shock adiabat existing at the 
the boundary between the mixed phase and quark gluon plasma. 

Thus,  all these new irregularities and signals discussed above are thoroughly verified on the existing experimental data and on the chemical FO parameters extracted from the data with the help of the HRGM. Therefore, they make a coherent picture and  give us  a strong evidence for the deconfinement occurrence at  $ \sqrt{s_{NN}}=$ 4.3--4.9 GeV.

\section{Conclusions} 

Here we critically  discussed the possible  signals of deconfinement PT outlined in the NICA White Paper 10.1 and showed that none of these signals is, in fact, worked out yet. Therefore, the weak point of the NICA White Paper is the absence of concrete suggestions what and where to measure in the future experiments to locate the mixed phase.

On the other hand,  above we presented a list of concrete observables whose peculiar behavior can serve as the solid foundation for  further  experimental verification. In fact, the results found recently 
in \cite{Oliinychenko:12,HRGM:13,KAB:gs, BWGwidth, SA1}, is a real breakthrough 
in the problem of formulating the realistic signals of the 1-sf order deconfining transition. 
They show that the future experiments have a chance to locate the mixed phase, if their accuracy is about an order magnitude better than now, and, if the energy scan steps are about 100 MeV in the laboratory frame. 

It is hoped, that the NICA and FAIR  Projects together with the BES program at RHIC and  the experiments performed at SPS CERN  will be able  to discover the mixed phase  and to experimentally locate the (tri)CEP of QCD phase diagram.  It seems,  however,  that 
without  developing 
a solid theoretical back up it will be too hard or even impossible to  achieve the declared  goals.

\vspace*{4.4mm}

{\bf Acknowledgments.}
The author  is   thankful to A. I. Ivanytskyi, D. H. Rischke, E. V. Shuryak, V. D. Toneev  and G. M. Zinovjev
for  fruitful discussions.  
Important  comments  of   D. B. Blaschke and  A. S. Sorin
are  acknowledged.
This publication is based on the research provided by the grant support of the State Fund for Fundamental Research   (project N  {\cyrfnt F}58/175-2014).

%%%\vspace*{2.2cm}
%%%KAB

\end{document}